\documentclass[twocolumn,prd,floatfix]{revtex4}
\usepackage{color}
\usepackage{tabularx}
\usepackage{epsfig}
\usepackage{amsmath}
\usepackage{amssymb}
\usepackage{bm}
\usepackage{graphicx}
\usepackage{multirow}

\usepackage{epsfig}

\begin{document}

\title{The Frequency-Shift in the Gravitational Microlensing}

\author{Sohrab Rahvar}
\affiliation {Department of Physics, Sharif University of Technology, Tehran 11155-9161, Iran 
\\ and \\ Department of Physics, College of Science, Sultan Qaboos University, P.O. Box 36, P.C. 123, Muscat, Sultanate of Oman
}

\date{\today}

\begin{abstract}

The relative transverse velocity of a lens with respect to the source star in gravitational lensing results in a frequency shift in the light rays passing by a lens. We propose using this relativistic effect for measuring the relative velocity of the lens with respect to the source star in gravitational microlensing. High precision spectrographs with the accuracy of detecting the relative frequency shift in the order of $10^{-11}$ will enable us to measure this effect in the microlensing events. The spectrographs such as ESPRESSO is going to be used for detecting exoplanets with the accuracy of the radial velocity of $0.1$~m/s. This kind of instrument can be used in follow-up observations of the microlensing events. Combining the spectroscopic observation with the parallax measurements of microlensing events from space and proper motion of the source stars with GAIA telescope enables us to measure all the parameters of the microlensing events. The result would be measuring the mass and the transverse velocity of lenses with the masses in the range of the black holes to the free-floating planets. 
\end{abstract}

\maketitle

The gravitational microlensing is the lensing of stars inside the Milky Way or nearby galaxies with the intervening compact objects such as other stars, brown dwarfs, black holes, Neutron stars or even planets. This class of gravitational lensing has become one of the astrophysical tools since 1986, following the suggestion of Paczysnki for detecting Massive Astrophysical Compact Halo Objects (MACHOs) in the halo of Milky Way galaxy \cite{pac}. The observations of microlensing events towards the Large and Small Magellanic Clouds excluded MACHOs as the candidate for the dark matter within the mass range of $(6\times 10^{-7},15)$ solar mass \cite{Tisserand}.  

The physical observable parameter in gravitational microlensing which is obtained from the light curve of each event is the Einstein crossing time. It depends on the physical parameters of the lens as   
\begin{eqnarray}
t_E &=& 45.6 day \left(\frac{D_s}{8.5 kpc}\right)^{1/2}\left(\frac{M}{0.5 M_\odot}\right)^{1/2}
\left( x(1-x) \right)^{1/2}\nonumber \\
&\times& \left(|{\bf v_{E,\bot} -v_{S,\bot}} - \frac{1}{x}({\bf v_{E,\bot} -v_{L,\bot}})|\frac{1}{220 km/s}\right)^{-1}
\label{te}
\end{eqnarray}
Here, $D_s$ is the distance of source and in the direction of the center of galaxy, we can set it  $D_s\sim 8.5$~ kpc, $M$ is the mass of lens, $x=D_l/D_s$ is the ratio of  the lens-distance to the source-distance from the observer, ${\bf v_{E\bot}}$,  ${\bf v_{S\bot}}$ and  ${\bf v_{L\bot}}$ are 
 the transverse relative velocities of the Earth, source and the lens in the Galactic frame \cite{MLbook}. 
  In equation (\ref{te}), measuring $t_E$,  we can not derive the mass, the distance and the transverse velocities of the lens and source. This is so-called the degeneracy problem in gravitational microlensing. In order to break the degeneracy between the lensing parameters, the perturbation effects such as parallax and finite-size effects deviate the simple microlensing light curve and provide the possibility to break partially this degeneracy \cite{rahvar,lee}.

In recent years, gravitational microlensing has been used for detection of the extra-solar planets. In this case, a planet orbiting around a parent star makes a binary lens and the effect of planet is a short-time scale perturbation in the microlensing light curve \cite{gaudi}. The observations of microlensing events both from the Earth and from space, such as Kepler \cite{kepler} and Spitzer \cite{spitzer} telescopes could provide a new opportunity to break the degeneracy between the lens parameters.  The observations of microlensing events with GAIA is also another possibility to break the degeneracy between the lens parameters by the astrometry of microlensing events \cite{dong}. 

In this work, we propose the spectroscopic observations of microlensing events for measuring the relative transverse velocity of the lens with respect to the source star. The transverse motion of a lens during gravitational lensing can change the energy of photons passing by the lens from a distant source. This effect for massive particles happens in the interplanetary travels of spacecrafts, so-called gravity-assist mechanism. There is a full general relativistic approach for the energy change of photons in \cite{pyne}. This effect has been studied for detection of the cosmic strings through the effect of relative motion of the cosmic strings with respect to the cosmic microwave background radiation \cite{smooth}. Also, the effect of a moving lens on the correction to the deflection angle is studied in \cite{kay,ser,Hey}.

\begin{figure}
\includegraphics[scale=0.55]{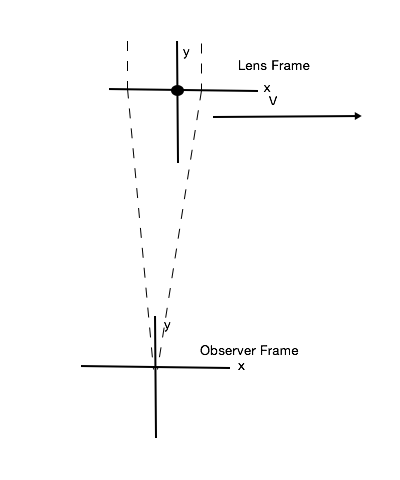}
\vspace{-1cm}
\caption{The relative transverse velocity of the lens with respect to the source star ($S$ frame) and the light deflection by the lens. }
\label{fig1}
\end{figure}

Here we use the coordinate transformation between the static and relative moving frames to 
derive the frequency change of photons after gravitational scattering and apply it to the gravitational microlensing.
Figure (\ref{fig1}) represents the relative transverse velocity of the lens with respect to a static observer. Here, the light rays arrive at the lens from a distant source and after bending receives by the observer. In our study, we define the  Minkowski reference frame ($S$) attached to the source star where photons are moving in the $y$ direction with the four energy-momentum components of $p^\mu = (p,0,p,0)$.  For photons, we adapt $E=p$, $p_y = p$ and set $c=1$. For the lens frame $(S^\prime)$ moving with the velocity of $({\bf v_L} - {\bf v_S})_i = v_{LS}\hat{i}$ as the relative velocity of lens with respect to the source star, the components of photon from the Lorentz transformation before the gravitational scattering, from the point of view of the lens frame is given by 
\begin{eqnarray}
\label{s0}
p^\prime_x(-\infty) &=& -\gamma v_{LS}E, \\
p^\prime_y(-\infty) &=& p_y,\nonumber \\
E^\prime(-\infty) &=& \gamma E.\nonumber 
\end{eqnarray}
We adapt "$-\infty$" notation from the particle physics convention, which means photons before the gravitational scattering at a long distance from the lens.

\begin{figure}
\vspace{-1.3cm}
\includegraphics[scale=0.65]{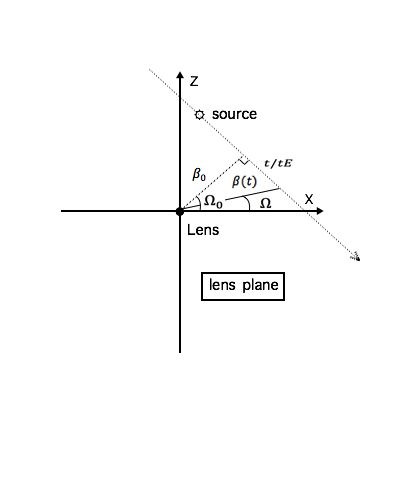}
\vspace{-3.0cm}
\caption{The relative transverse motion of the source (depicted by dotted line) with respect to the lens on the lens plane.
The lens is located at the center of coordinate system and the observer is along the y-axis. The angle of $\Omega$ is the relative angle between the position of the source and $x$ axis.  $\Omega_0$ is defining this angle at minimum impact parameter (i.e. $\beta (t = 0) = \beta_0$) and $t/t_E$ is the displacement of the source from the position of the minimum impact parameter.}
\label{fig2}
\end{figure}

Let us assume a flat-plane going through the observer, lens, and source. This plane is so-called the lensing plane and photos that are received by the observer are going on this plane. In general, this plane has an angle with the $x-y$ plane where the cross-section of it with the $x-z$ plane produces the angel of $\Omega$ with respect to the $x$-axis (see Figure \ref{fig2}).  
According to the light bending in Schwarzschild metric the deflection angle is give by $\alpha = 4GM/b$ where $M$ is the mass of lens and $b$ is the closest distance of the light ray to the lens.  We can represent $\alpha$ as a vector in $x-z$ plane as $ \vec{\alpha} =- \alpha\cos\Omega \hat{i}-  \alpha\sin\Omega \hat{k}$ where the minus sign represents that deflection is toward the lens. We replace the deflection angle with $\vec{\Delta p\prime}= \vec{\alpha} p_y$ on $x-z$ plane where $\vec{\Delta p\prime}$ is the extra momentum of photons that is gaining during the lensing with the following components of  $\Delta p_x^\prime = -\alpha p_y\cos\Omega$ and  $\Delta p_z^\prime = -\alpha p_y\sin\Omega$.

After scattering of photons from the lens, we add the extra components of momentum of photons from the lensing. So the new energy-momentum of photons from the set of equations (\ref{s0}) after gravitational scattering 
can be written as 
\begin{eqnarray}
\label{s1}
p^\prime_x &=& -\gamma v_{LS}E - \alpha p_y \cos\Omega,   \\
p'_y &=& p_y, \nonumber \\
p^\prime_z &=& - \alpha p_y \sin\Omega,  \nonumber \\
E^\prime &=& \gamma E. \nonumber
\end{eqnarray}
Finally we again return back to the $S$ frame at the distance of $y\rightarrow +\infty$. In this case, the frame is moving with  $\vec{v} = -v_{LS}\hat{i}$ with respect to the lens (e.g. $S'$).  In the third frame $(S'' = S)$, the energy-momentum components of photons after Lorentz transformation is related to the $S'$ frame as 
\begin{eqnarray}
\label{s2}
p''_x(+\infty) &=& \gamma(p'_x + v_{LS} E'), \\
p''_y(+\infty)  &=& p'_y, \nonumber \\
p''_z(+\infty) &=& p'_z, \nonumber \\
E''(+\infty)    &=& \gamma (E'+v_{LS} p'_x). \nonumber 
\end{eqnarray}
Substituting the set of equations of (\ref{s1}) in equations (\ref{s2}) results in the energy-momentum of the photons with respect to the distant observer in $S$ frame. 
\begin{eqnarray}
\label{s3}
p''_x (+\infty) &=& -\gamma\alpha p_y\cos\Omega \\
p''_y(+\infty)  &=& p_y, \nonumber \\
p''_z(+\infty) &=& -\alpha p_y \sin\Omega, \nonumber \\
E''(+\infty) &=& E(-\infty)(1 - \gamma v_{LS} \alpha \cos\Omega) \nonumber 
\end{eqnarray}
where we used $p_y = E(-\infty)$. 

According to the components of photon's momentum in three dimension, the gravitational deflection 
angle is $\vec{\alpha} = \Delta \vec{p}/p_y = -\alpha (\gamma\cos\Omega~\hat{i} + \sin\Omega~\hat{k})$ where for the case of $\gamma=1$, we recover the standard Schwarzschild deflection angle from the gravitational lensing. The deflection angle in $x$-direction has the extra correction factor of $\gamma$ which implies a correction in the order of $\mathcal{O}(v^2)$ compare to the static lens. On the other hand, the energy change from equation (\ref{s3}) due to the transverse velocity of lens is $\Delta E/E = -\gamma v_{LS} \alpha \cos\Omega$ where keeping  $\mathcal{O}(v)$  terms, the frequency change results in 
\begin{equation}
\label{nu}
\frac{\delta\nu}{\nu} = - \frac{4GM}{b} v_{LS}  \cos\Omega.
\end{equation}
 The frequency shift equation can also be obtained from the Shapiro delay formalism \cite{afshordi}. An extensive investigation on gravitational lensing of moving lenses has been studied in \cite{sergey}. Similar to the gravity assist in terrestrial mechanics, photons that passing in front of the lens (i.e $|\Omega|<\pi/2$) loos energy while photon passing behind the lens (i.e $\pi/2<|\Omega|<\pi|$) gains energy. If we have a point-like lens and two images from a single source, one of them should be blue-shifted and the other one red-shifted. The amount of this frequency change depends on the transverse velocity of the lens as well as the deflection angle. For an observer located at far distance from the lens and having a relative velocity with respect to the source star, there is an additional correction in the order of $\mathcal{O}((v_{E} - v_{S})^2)$ which can be ignored compare to the $\mathcal{O}(v_{LS})$ in equation (\ref{nu}).  So the frequency shift to the observer on Earth is given by equation (\ref{nu}). 

In gravitational microlensing, the observer, the lens and the source star have a relative motion and the lensing configuration changes by time. So, the lensing plane which contains the observer-lens-source, as well as the trajectory of light, will change by time with respect to the X-axis (representing the relative velocity of the lens with respect to the source).  In Figure (\ref{fig2}), the relative motion of the source with respect to the lens is plotted by the dotted line on the lens plane. We use $\beta(t)$ as the impact parameter (position of the source in absent of lens) normalized to the Einstein angle, (i.e. $\theta_E =\sqrt{4GM(D_s-D_l)/D_sD_l} $), $\beta_0$ is the minimum impact parameter and $\Omega_0$ is the relative angle for the position of the source to the lens on the lens plane with respect to the $X$-axis at the minimum impact parameter.  

We note that in the simple lensing (point mass source and lens), the two images form on the lens plane along the position of the source with respect to the lens where the positive parity image forms with the angle of $\Omega$ compare to the X-axis and the negative parity image forms at the angle of $\Omega + \pi$.  Using the geometry of the relative trajectory of the source on the lens plane (in Figure (\ref{fig2})), $\Omega$ as a function of time is given by 
\begin{equation}
\label{omega}
\cos\Omega= \frac{t/t_E\times\sin\Omega_0 + \beta_0\cos\Omega_0}{\beta(t)}, 
\end{equation}
where $t$ is the time and $t=0$ is the moment where the position of the source is at the minimum impact parameter, (i.e. $\beta(t=0) = \beta_0$). Substituting equation (\ref{omega}) into (\ref{nu}) results in the relative frequency change as
\begin{eqnarray}
\label{dnu}
\frac{\delta\nu}{\nu} &=& -\frac{2R_S}{R_E}\times v_{LS}\times\frac{t/t_E\sin\Omega_0 + \beta_0 \cos\Omega_0}{\theta(t)\beta(t)}.\nonumber \\
&=&- \frac{C(t)}{\theta(t)},
\end{eqnarray}
where $R_S$ and $R_E$ are the Schwarzschild radius the Einstein radius of the lens and $C(t)$ is
\begin{eqnarray}
&& C(t) = 2.23\times 10^{-12} \left(\frac{M}{0.5 M_\odot}\right)^{1/2}\left(\frac{D_s}{8.5~\text{kpc}}\right)^{-1/2}\nonumber \\
&\times& \left(\frac{v_{LS}}{200~ \text{km/s}}\right) \left(x(1-x)\right)^{-1/2} \frac{t/t_E\sin\Omega_0 + \beta_0\cos\Omega_0}{\beta(t)}. \nonumber\\
&&
\label{fshift}
\end{eqnarray}
Here we assume that the source stars for gravitational microlensing events are located at the center of galaxy $D_s = 8.5~$kpc with the mass of lens in the order of $M = 0.5 M_\odot$ and a relative velocity of the lens-source of $v_{LS} = 200$ km/s. This condition happens if the source star resides in the Galactic bulge and the lens is located at the Galactic disk. Stars in the Galactic bulge have almost isotropic dispersion velocity of $100$ km/s while stars in the disk have the dominant global velocity of $\sim 200$ km/s with a small dispersion velocity in the order of $10$ km/s \cite{binney}.  
 The positive and negative parity images from the lensing equation result in the opposite frequency shifts where $C(t)$ for both images are the same and $\theta(t)$ is different for the two images.  Assuming a point-like source, the solution of the lens equation provides the transformation between the source position to the image position for the two positive and negative parity images as follows:
\begin{equation}
\label{le}
\theta^\pm = \frac{1}{2}(\beta \pm \sqrt{\beta^2 + 4}),
\end{equation}
 where all the angles are normalized to the Einstein angle.
If we take a Dirac-delta function as the spectral line with the frequency of $\nu_0$, we expect that in the positive and the negative parity images frequencies change as follows
\begin{equation}
\nu_0^\pm = \nu_0(1-\frac{C}{\theta^\pm}).
\end{equation}

The magnification for the positive and negative parity images is also given by 
\begin{equation}
|A^\pm |= |\frac{{\theta^\pm}}{\beta}\frac{\partial\theta^\pm}{\partial\beta}|,
\end{equation}
where substituting equation (\ref{le}) in the magnification results in 
\begin{equation} 
|A^\pm | = \frac{1}{2}|1\pm \frac{2+\beta^2}{\beta\sqrt{\beta^2 + 4}}|, 
\label{amp}
\end{equation}
and the overall magnification is 
\begin{equation}
A = \frac{2+\beta^2}{\beta\sqrt{4 + \beta^2}}.
\end{equation}
For the case of extended sources, we should consider the finite-size effect \cite{rahvar,gholchin} in our calculations.

 We note that in gravitational microlensing, the configuration of lens with respect to the source changes with time  (as shown in Figure \ref{fig2}) and as a result, the spectral pattern also changes with time.
 In order to study the time variation of the spectral line, we can define the relative overall frequency shift in the light curve weighted by the flux of each images to measure the overall shift of the spectral barycenter, as follows:
\begin{equation}
\label{dnut}
\frac{\delta\nu}{\nu}  = \frac{1}{A}\left(\frac{\delta\nu^{+}}{\nu}|A^{+}| + \frac{\delta\nu^{-}}{\nu}|A^{-}|\right),
\end{equation}
where $A = |A^+| + |A^-|$ is the total magnification from the lensing. By substituting equations (\ref{dnu}) and (\ref{amp}) in (\ref{dnut}), the frequency shift from the two images obtain as follows:
\begin{equation}
\label{dnut2}
\frac{\delta\nu}{\nu}  = - \frac{C(t)}{A}\left(\frac{|A^{+}|}{|\theta^+|} - \frac{|A^{-}|}{|\theta^-|}\right),
\end{equation}
where for a dynamical impact parameter, $\beta^2(t) = \beta_0^2 + (t/t_E)^2$, the overall frequency 
shift as a function of time is given by 
\begin{eqnarray}
&& \frac{\delta\nu}{\nu} = -2.23\times 10^{-12} (\frac{M}{0.5 M_\odot})^{1/2}(\frac{D_s}{8.5kpc})^{-1/2}\nonumber \\
&\times& (\frac{v_{LS}}{200 km/s}) \left(x(1-x)\right)^{-1/2} \frac{t/t_E\sin\Omega_0 + \beta_0\cos\Omega_0}{2+\beta_0^2+(t/t_E)^2}. \nonumber\\
&&
\label{fshift}
\end{eqnarray}

\begin{figure}
\includegraphics[scale=0.45]{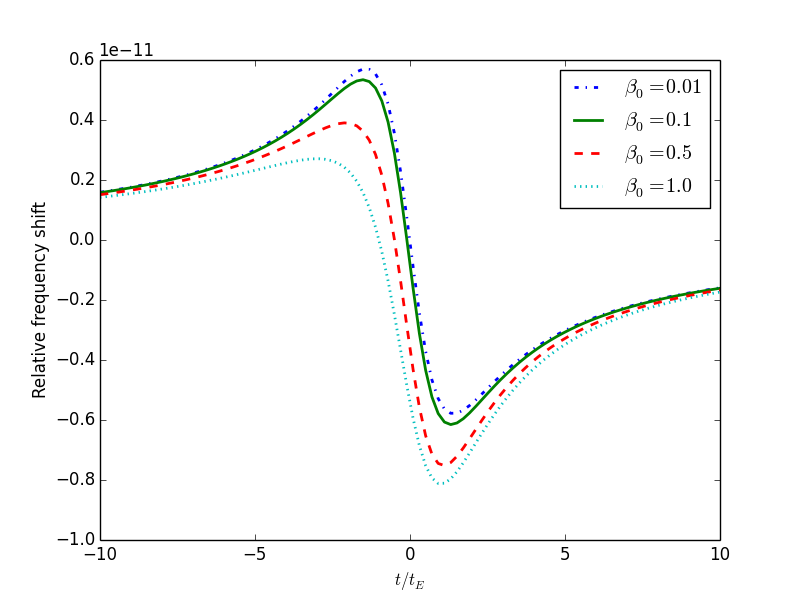}
\caption{The frequency shift as a function of time (normalized to $t_E$) for a lens with the typical values adapted from equation (\ref{fshift}). The frequency shift is plotted for four minimum impact parameters.}
\label{fig3}
\end{figure}

Figure (\ref{fig3}) shows the frequency shift of a spectral line in the microlensing event as a function of time, normalized to the Einstein crossing time, for different values of minimum impact parameters. Here, we adapt $x=0.1$ and lens mass is $M=0.5 M_\odot$.  The frequency shift for different values of $\beta_0$ is about $\sim 10^{-11}$. The frequency shift due to transverse velocity of lens also has been detected using radio links with the Cassini spacecraft \cite{casini} and 
the relative frequency shift from the observation is similar to Figure (\ref{fig3}) of this work.

 Let us compare the frequency shift from equation (\ref{fshift}) with the resolution of recent spectrographs such as ESPRESSO (Echelle SPectrograph for Rocky Exoplanets and Stable Spectroscopic Observations) \footnote{https://www.eso.org/sci/facilities/paranal/instruments\\ /espresso.html} for detection of radial velocity in exoplanet observations \cite{spect,espresso}. This spectrograph in high resolution mode within the wavelength range of $[380,778]$~nm and the limiting magnitude of $m<17$ can measure the radial velocity of the order of $v=0.1$ m/s  which corresponds to the frequency shift in the spectral lines of $\delta \nu/\nu =v/c\simeq  3\times 10^{-10}$. While the ESPRESSO resolution is lower than what we expect for the microlensing observation, however for the massive lenses we expect detection of this effect.

Summarizing this work, the advance spectrographs can perform spectroscopy of the on-going microlensing 
events for measuring the frequency shift in the spectrum of the source stars. This effect results from 
the relative transverse velocity of lens with respect to the source star and measuring this effect provides 
$v_S-v_L$ as one of the unknown parameters of the microlensing events. On the other hand astrometry of 
the source stars in the microlensing events with GAIA can provide $v_S-v_E$. Combining these two equations, 
we can provide the relative velocity of source and lens with respect to the earth. 
On the other hand, recent parallax observations of microlensing events by space-based telescopes \cite{zhang} 
can put extra constrain between the mass and the distance of the lens. We can extract all the parameters 
of the lens by combining the space-based parallax observations and the frequency shift of the source 
star (we have introduced in this work).  These network of observations will enable us to discover the 
mass and velocity of all kind of dark compact objects (as the lens of microlensing events) from the 
astrophysical black holes to the free-floating planets \cite{sumi}.

I would like to thank referee you his/her useful comment for improving this paper. 
This research was supported by Sharif University of Technology’s Office of Vice 
President for Research under Grant No. G950214,

\bibliographystyle{apsrev}

\begin{thebibliography}{99}


\bibitem{pac} 
B. Paczynski, ApJ, {\bf 304},  1 (1986) 

\bibitem{Tisserand}
P. Tisserand et al. A\&A {\bf 469}, 387 (2007)

\bibitem{MLbook}
S. Mollerach, E. Roulet, Gravitational Lensing and Microelsning, World Scientific (2002)


\bibitem{rahvar}
S. Rahvar, Int. J. of Mod. Phys. D {\bf 24}, 1530020 (2015)

\bibitem{lee}
Chien-Hsiu Lee, Universe 2017, 3, 53

\bibitem{gaudi}
B. Scott Gaudi, Annual Review of Astronomy and Astrophysics {\bf 50},411 (2012) 

\bibitem{kepler}
C. B. Henderson et al. Publications of the Astronomical Society of the Pacific {\bf 128}, 970 (2016)

\bibitem{spitzer}
R. Poleski, W. Zhu, G.~W. Christie et al. ApJ {\bf 823}, 63 (2016)

\bibitem{dong}
S. Dong et al. ApJ {\bf 871}, 70  (2019)

\bibitem{pyne}
T. Pyne, M. Birkinshaw, ApJ {\bf 415}, 459 (1993)

\bibitem{smooth}
E. Jeong, C. Baccigalupi, G.F. Smoot, JCAP {\bf 09}, 018 (2010)


\bibitem{kay}
R. Kayser, S. Refsdal, R.Stabell, Astronomy and Astrophysics, vol. 166, no. 1-2,
1986, p. 36-52 (1986)  

\bibitem{ser}
M. Sereno, Mon. Not. R. Astron. Soc. 359, L19-L22 (2005)

\bibitem{Hey}
D. Heyrovsky, The Astrophysical Journal, 624, 28-33 (2005)

\bibitem{afshordi}
Sh. Baghram, N.  Afshordi, K. M. Zurek, Physical Review D  84, 4, 043511 (2011)

\bibitem{sergey}
S. M. Kopeikin, G. Schafer, Physical Review D 60, 12, 124002 (1999)
 

\bibitem{binney}
J. Binney, S. Tremaine, Galactic dynamics, (1987) 

\bibitem{gholchin}
L. Golchin, S. Rahvar, arXiv:1906.10589  (2019)

\bibitem{casini}
B. Bertotti, L. Iess \& P. Tortora, Nature 425, 374 (2003) 



\bibitem{spect}
F. Pepe, et al. Astron. Nachr., 335: 8-20 (2014)


\bibitem{espresso}
Jonay I. González Hernández, et al, Handbook of Exoplanets, ISBN 978-3-319-55332-0. Springer International Publishing AG, part of Springer Nature, 2018, id.157;  arXiv:1711.05250


\bibitem{zhang}
W. Zang et al. , arXiv:1904.11204 (2019)

\bibitem{sumi}

T. Sumi, et al. Nature 473, 349 (2011)


\end{thebibliography}

\end{document}